\documentclass[epj]{svjour}
\usepackage{latexsym}
\usepackage{amsmath,amssymb}
\usepackage{graphics}
\begin{document}
\title{Studies of the Giant Dipole Resonance in $^{27}$Al, $^{40}$Ca, $^{56}$Fe, $^{58}$Ni and $^{208}$Pb with high energy-resolution inelastic proton scattering under 0$^\circ$}

\titlerunning{Studies of the Giant Dipole Resonance with high energy-resolution inelastic proton scattering}

\author{M.~Jingo \inst{1,2} \and E.Z. Buthelezi\inst{3} \and J.~Carter\inst{1} \and  G.R.J.~Cooper\inst{4} \and R.W.~Fearick\inst{5} \and S.V.~F\"{o}rtsch\inst{3}\and C.O.~Kureba\inst{1} \and A.M.~Krumbholz\inst{6} \and P.~von~Neumann-Cosel\inst{6} \and R.~Neveling\inst{3}  \and P.~Papka\inst{7} \and I.~Poltoratska\inst{6} \and V.Yu.~Ponomarev\inst{6} \and A.~Richter\inst{6} \and E.~Sideras-Haddad\inst{1} \and F.D.~Smit\inst{3} \and J.A.~Swartz\inst{7} \and A.~Tamii\inst{8} \and I.T.~Usman\inst{1}}
%
%
\institute{School of Physics, University of the Witwatersrand, Johannesburg 2050, South Africa \and Department of Orthopaedic Surgery, University of the Witwatersrand, Johannesburg 2193, South Africa \and Department of Subatomic Physics, iThemba LABS, PO Box 722, Somerset West 7129, South Africa \and School of Earth Sciences, University of the Witwatersrand, Johannesburg 2050, South Africa \and Department of Physics, University of Cape Town, Rondebosch 7700, South Africa \and Institut f\"{u}r Kernphysik, Technische Universit\"{a}t Darmstadt, D-64289 Darmstadt, Germany \and Department of Physics, University of Stellenbosch, Matieland 7602, South Africa \and Research Center for Nuclear Physics, Osaka University, Ibaraki, Osaka, 567-0047, Japan}
\date{Received: date / Revised version: date}
%
\abstract{
A survey of the fine structure of the Isovector Giant Dipole Resonance (IVGDR) was performed, using the recently commissioned Zero-degree Facility of the K600 magnetic spectrometer at iThemba LABS. 
Inelastic proton scattering at an incident energy of 200 MeV was measured on $^{27}$Al, $^{40}$Ca, $^{56}$Fe, $^{58}$Ni and $^{208}$Pb. 
A high energy-resolution ($\rm{\Delta}\it{E} \simeq$ 40 keV FWHM) could be achieved after utilising faint-beam and dispersion-matching techniques. 
Considerable fine structure is observed in the energy region of the IVGDR and characteristic energy scales are extracted from the experimental data by means of a wavelet analysis. 
The comparison with Quasiparticle-Phonon Model (QPM) calculations provides insight into the relevance of different giant resonance decay mechanisms.   
Photoabsorption cross sections derived from the data assuming dominance of relativistic Coulomb excitation are in fair agreement with previous work using real photons. 
} 
\maketitle

\section{Introduction}
\label{intro}

Spectroscopy of charged-particle induced reactions provides a potentially rich source of nuclear structure information, and extensive studies have been performed e.g.\ with electrons and protons up to heavy ions as probes. 
The use of magnetic spectrometers in nuclear structure studies brought tremendous improvement in the analysis of nuclear reaction products due to their high resolution capabilities. 
Spectroscopy with magnetic spectrometers offers several advantages compared to other detection techniques using gas-ionisation, semi-conductors or scintillation detectors. 
Among these is the strong selection of the reaction products based on their mass-over-charge ratio which results in the reduction of background in the spectra. 
This enables the detection of the reaction products at very forward scattering angles with magnetic spectrometers.

The interest in studies at $0^\circ$ is triggered by the extreme selectivity of the reactions to low angular momentum transfer. 
As an example, using a combination of 0$^\circ$ cross sections with high energy-resolution, a detailed understanding of the Gamow-Teller (GT) mode in nuclei has been reached \cite{fuj11}. 
At the  the Research Center for Nuclear Physics (RCNP) in Osaka, Japan, the ($^{3}$He,t) reaction was established as a tool for determination of GT$_{-}$ (isospin \textit{T} = $T_{\rm0} - 1$, where $T_{\rm0}$ is the ground state isospin of the target nucleus) reaching typical energy resolutions $\rm{\Delta}\it{E \approx}$ 30 keV (FWHM) in light- \cite{sch11} and $\approx$ 50 keV in heavy-mass nuclei \cite{kal06}. 
At the Kernfysisch Versneller Instituut (KVI), Groningen, Netherlands, similar success was also achieved with the (d,$^{2}$He) reaction as a measure of GT$_{+}$ strength from (n,p)-type reactions \cite{hag04}. 

Inelastic proton scattering has been investigated as a probe in the early days of giant resonance studies (see, e.g., refs.~\cite{lew72,ber79} and references therein). 
This work contributed to establishing the systematics of the Isoscalar Giant Monopole Resonance (ISGMR) and Isoscalar Giant Quadrupole Resonace (ISGQR), but later $\alpha$ scattering was preferred for the investigation of these modes because of its selectivity to isoscalar excitations. 
Recently, inelastic proton scattering at $0^\circ$ and incident beam energies of a few hundred MeV has been used to probe the IVGDR utilizing the dominance of Coulomb excitation in these kinematics.
However, such an experiment presents an experimental challenge due to the very small magnetic rigidity difference between the incident and scattered particles.
Consequently, these measurements are known to be extremely sensitive to beam halo and background from atomic small-angle scattering in the target since the primary beam exits the spectrometer very close to the position of the focal plane detectors. 
Historically, (p,p$'$) measurements under $0^\circ$ at intermediate beam energies (100 -- 300 MeV) with reasonably low background have been successfully carried out only at two places, namely the RCNP \cite{tam09} and the Indiana University Cyclotron Facility (IUCF) in Bloomington, Indiana, USA \cite{IUCF}. 
However, such measurements can now also be carried out at the iThemba Laboratory for Accelerator Based Sciences (iThemba LABS) near Cape Town, South Africa \cite{nev11}.
Here, we report the results of a proof-of-principle study of $0^\circ$ (p,p$'$) scattering at iThemba LABS covering nuclei over a wide mass range performed to assess the capabilities and limitations of the newly commissioned Zero-degree Facility.

Although originally motivated by the possibility to study the spin-$M$1 strength in nuclei  \cite{hey10,iwa12,mat15,bir16,mat17}, an important aspect of carrying out experiments such as the ones reported here is the possibility to extract the complete electric dipole ($E$1) strength distribution (excited via relativistic Coulomb excitation) from excitation energies starting well below neutron threshold across the excitation region of the IVGDR \cite{tam11,pol12,kru15,has15,bir17,mar17}.
This allows an extraction of the nuclear dipole polarizability, which provides important information on the neutron skin thickness of nuclei and the density dependence of the symmetry energy \cite{roc18}.
 
The power of forward-angle inelastic proton scattering at energies of several 100 MeV, in particular if combined with dispersion-matching techniques enabling high energy-resolution, was demonstrated for the reference case of $^{208}$Pb \cite{tam11,pol12}. 
In this case, the $E$1 strength is well known through studies of the $^{208}$Pb($\gamma$,$\gamma{'}$) \cite{rye02} and photoabsorption reactions \cite{vey70,sch88}.
Historically, dipole strengths were studied by combining data from different reactions such as ($\gamma$,xn) and ($\gamma$,$\gamma{'}$), or particle-$\gamma$ coincidence reactions such as ($^{3}$He,$^{3}$He$'\gamma$). 
However, ($\gamma$,xn) measurements are restricted to the energy region above the neutron-separation energy ($S_{\rm n}$), while ($\gamma$,$\gamma{'}$) reactions yield data for low excitation energies roughly up to the neutron threshold and potentially large corrections due to branching ratios to excited states have to be applied \cite{rus09}. 
However, in (p,p$'$) scattering at small scattering angles including 0$^{\circ}$, the investigation is independent of this problem and the $E$1 response can be probed both above and below the neutron-separation energy on the same footing.

Another interesting phenomenon that can be studied in such experiments is that of fine structure in the excitation region of the IVGDR. 
Pioneering work on the fine structure of giant resonances was reported a long time ago in high energy-resolution ($\rm{\Delta}\it{E \simeq}$ 50 keV) electron scattering experiments at the DALINAC \cite{sch75,kuh86}, where the Isoscalar Giant Quadrupole Resonance (ISGQR) in $^{208}$Pb was investigated. 
The same structures later seen in high resolution proton scattering experiments \cite{kam97} and
a survey over a wide nuclear mass-range demonstrated that the fine structure is not specific to $^{208}$Pb but generally appears in even-even nuclei \cite{she04,she09}. 
Moreover, fine structure has been observed in many other types of giant resonances such as the IVGDR \cite{str00,pol14}, the Magnetic Quadrupole Resonance \cite{vnc99} and the GT mode \cite{kal06}, establishing it as a general phenomenon in nuclei.

The nuclear giant resonance width $\rm \Gamma$ is determined mainly by three different mechanisms which can all cause fine structure: fragmentation of the elementary one particle-one hole (1p-1h) excitations with an average energy $\rm{\Delta}\it{E}$ (Landau damping), direct particle decay out of the continuum (escape width $\rm \Gamma\!\uparrow$), and statistical particle decay due to coupling to two (2p-2h) and many particle-many hole (np-nh) states (spreading width $\rm \Gamma\!\downarrow$)
\begin{equation}
\rm \Gamma = \rm{\Delta} \textit{E} + \Gamma\!\uparrow + \Gamma\!\downarrow.
\end{equation} 
Recently, it has been demonstrated that the wavelet method is a proper tool shown for the quantitative analysis of the fine structure observed in the spectra \cite{kal06,she04}. 
It was shown that the experimental observation of fine structure of the ISGQR in medium- and heavy-mass nuclei is related to the coupling between low-energy collective states and more complex two-particle two-hole (2p-2h) states, while in lighter-mass nuclei also Landau damping seems to be important \cite{usm11}. 
A recent study of the fine strucure of the IVGDR in $^{208}$Pb indicates that again Landau damping is most important \cite{pol14}.

One purpose of the present study is to extend the systematics of the fine structure in the case of the IVGDR, which may be considered the archetype of giant resonances in nuclei. 
This is made possible by the advent of the K600 magnetic spectrometer Zero-degree Facility of iThemba LABS using high energy-resolution inelastic proton scattering at $E_{\rm p}$ = 200 MeV. 
Due to beam line constraints, scattering angle discrimination is not possible at $0^{\circ}$, and measurements at finite angle and polarisation measurements are presently not available.
Thus a decomposition from other contributions expected in the data like the spin-flip $M$1 resonance, the ISGMR and the ISGQR  as performed e.g.\ in refs.~\cite{pol12,kru15} is not possible.
However, relativistic Coulomb excitation of non-splin-flip $E$1 transitions dominates the cross sections close to $0^\circ$.
A follow-up experiment on the chain of stable Nd isotopes \cite{don18} demonstrated that the
ISGMR and ISGQR cross sections are small (less than 10\% extrapolated to the lighter targets investigated in the present paper).
The spin-flip $M$1 resonance is well known in $sd$- and $fp$-shell nuclei \cite{hey10} and also in $^{208}$Pb \cite{bir16} and lies at excitation energies well below the IVGDR. 
The main background in the spectra stems from quasi-free scattering, which is approximated by a phenomenological approach as discussed below. 

Albeit being the best studied case experimentally and theoretically, many open questions remain in our understanding of the IVGDR \cite{har01}. 
In particular, the observed resonance widths as a function of mass number show strong fluctuations which have never been fully explained on a microscopic basis \cite{jun08}. 
An in-depth analysis of the fine structure can certainly add new and independent information to this problem and permits stringent tests of the available microscopic theoretical models. 
Therefore, a systematic study of light- to heavy-mass nuclei was undertaken.

\section{Experiments}
\label{sec:expt}

The experiments were carried out with a 200 MeV proton beam produced by the Separated Sector Cyclotron (SSC) at iThemba LABS. 
The protons were scattered inelastically off targets within a wide mass-range, namely  $^{27}$Al, $^{40}$Ca, $^{56}$Fe, $^{58}$Ni and $^{208}$Pb.
Highly enriched ($>95$\%) targets with typical areal densities of $1 - 3$ mg/cm$^{2}$ were used. 
In the case of $^{40}$Ca the target material consisted of natural calcium (97\% $^{40}$Ca). 
Scattered protons were momentum analysed with the K600 magnetic spectrometer, which was placed at $0^\circ$ and covered laboratory scattering angles $\theta_{\rm lab} \leq 1.91^{\circ}$.
The experimental kinematics favours excitation of dipole modes by relativistic Coulomb excitation ($E$1) and $\rm{\Delta}\it{L}$ = 0 spin-flip ($M$1) transitions \cite{tam11,pol12}. 
Dispersion-matching techniques were used in order to exploit the high energy-resolution capability of the spectrometer. 
Energy resolutions of $\rm{\Delta}\it{E \simeq} ~\rm 40 - 50$ keV (FWHM) were achieved. 
The focal plane detector system consisted of two Vertical Drift Chambers (VDCs) and a pair of rectangular plastic scintillation detectors behind them.
Details on the experimental setup and data reduction beyond the description provided in this section can be found in refs. \cite{nev11,jin14}.
\subsection{Particle identification}
\label{sec:Part}

\begin{figure}
\begin{center}
\resizebox{0.4\textwidth}{!}{%
\includegraphics{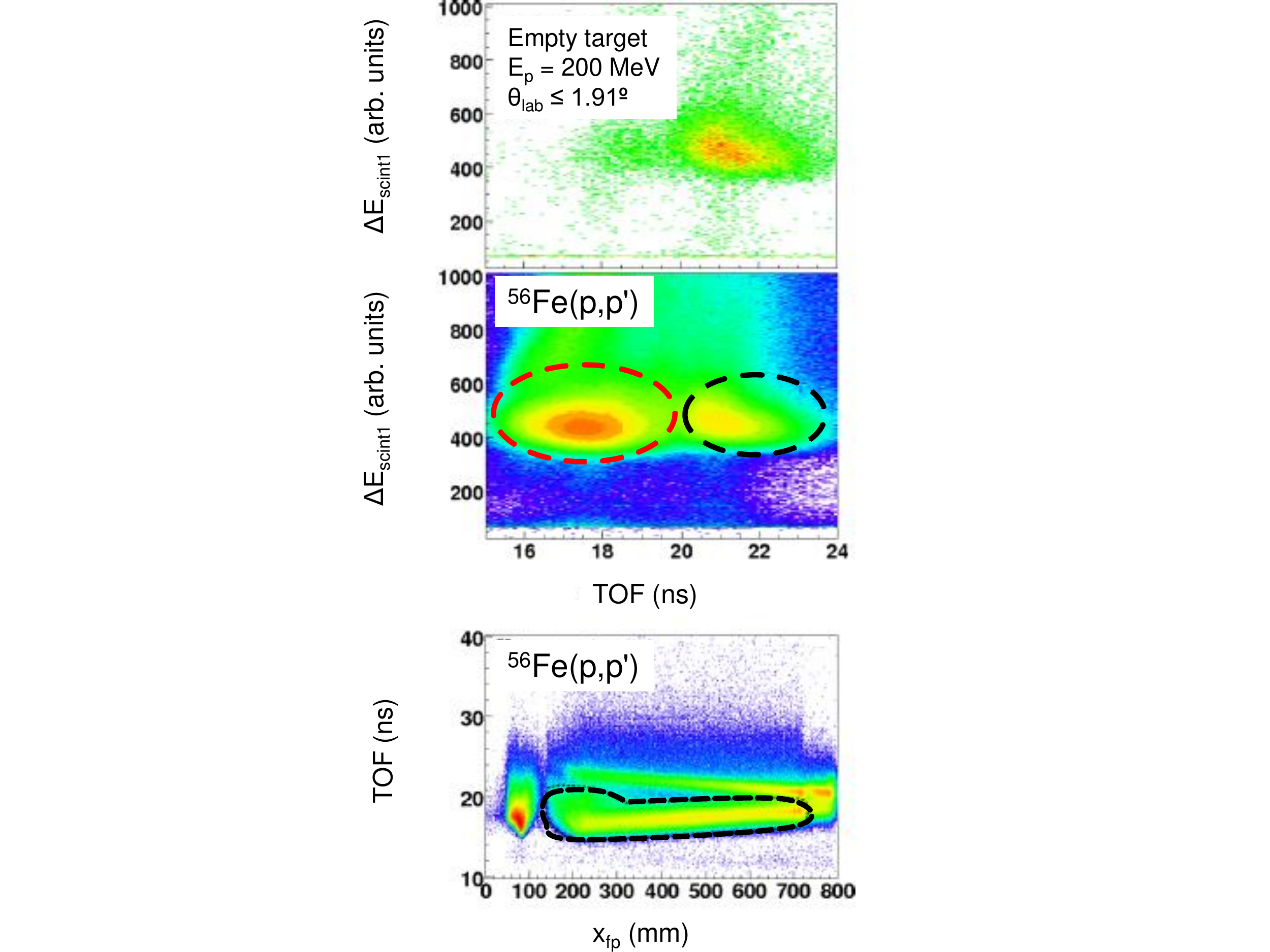}
}
\end{center}
\caption{Background in the (p,p$'$) reaction at $E_{\rm p} = 200$ MeV and $\theta_{\rm lab} = 0^\circ - 1.91^\circ$.
Top panel: Two-dimensional scatterplot showing the locus of the pulse height of the first scintillator  caused by beam halo \textit{versus} TOF. 
Middle panel: Same for a $^{56}$Fe target.
The right locus (dashed black-circle) represents beam halo events, and the left locus (dashed red-circle) represents protons scattered from the target.
Bottom panel: Two-dimensional scatterplot of TOF \textit{versus} horizontal focal position ($x_{\rm fp}$) showing loci of the inelastically scattered protons and beam halo as well as other instrumental background. 
The dashed black-curve represents the two-dimensional gate used in the further analysis to identify events due to the (p,p$'$) reaction.
\label{fig:pdtof}}
\end{figure}

Particle identification relies on the combination of knowledge of the accepted rigidity range in the focal plane (determined by the magnetic field), the energy loss of the particle in the plastic scintillators, as well as the Time of Flight (TOF) of the particle in the spectrometer.  
For $0^\circ$ inelastic proton scattering experiments PID techniques are especially important to help distinguish between protons scattered from the target and beam halo events.
The characteristics of the energy loss and TOF of events caused purely by beam halo can be determined by an empty target measurement. 
The result of such a measurement is shown in the scatterplot in the top panel of fig.~\ref{fig:pdtof}.  
The TOF is determinated by measuring the time elapsed between a coincident scintillator signal and the radio-frequency signal from the SSC, and as such represents a relative measurement of the TOF.  
 
The middle panel in fig.~\ref{fig:pdtof} illustrates the energy loss and TOF characteristics associated with events from a $^{56}$Fe target.
It is clear that the majority of background events can be distinguished from target-related events. 
The instrumental background can be suppressed by an appropriate software gate on the target related events. 
This can be further improved by selecting the appropriate events (i.e. inelastically scattered protons) in the scatterplot of TOF \textit{versus} focal plane position ($x_{\rm fp}$) as indicated in the bottom panel of fig.~\ref{fig:pdtof}.

\subsection{Subtraction of background}
\label{sec:bg}

\begin{figure*}
\begin{center}
\resizebox{0.75\textwidth}{!}{%
\includegraphics{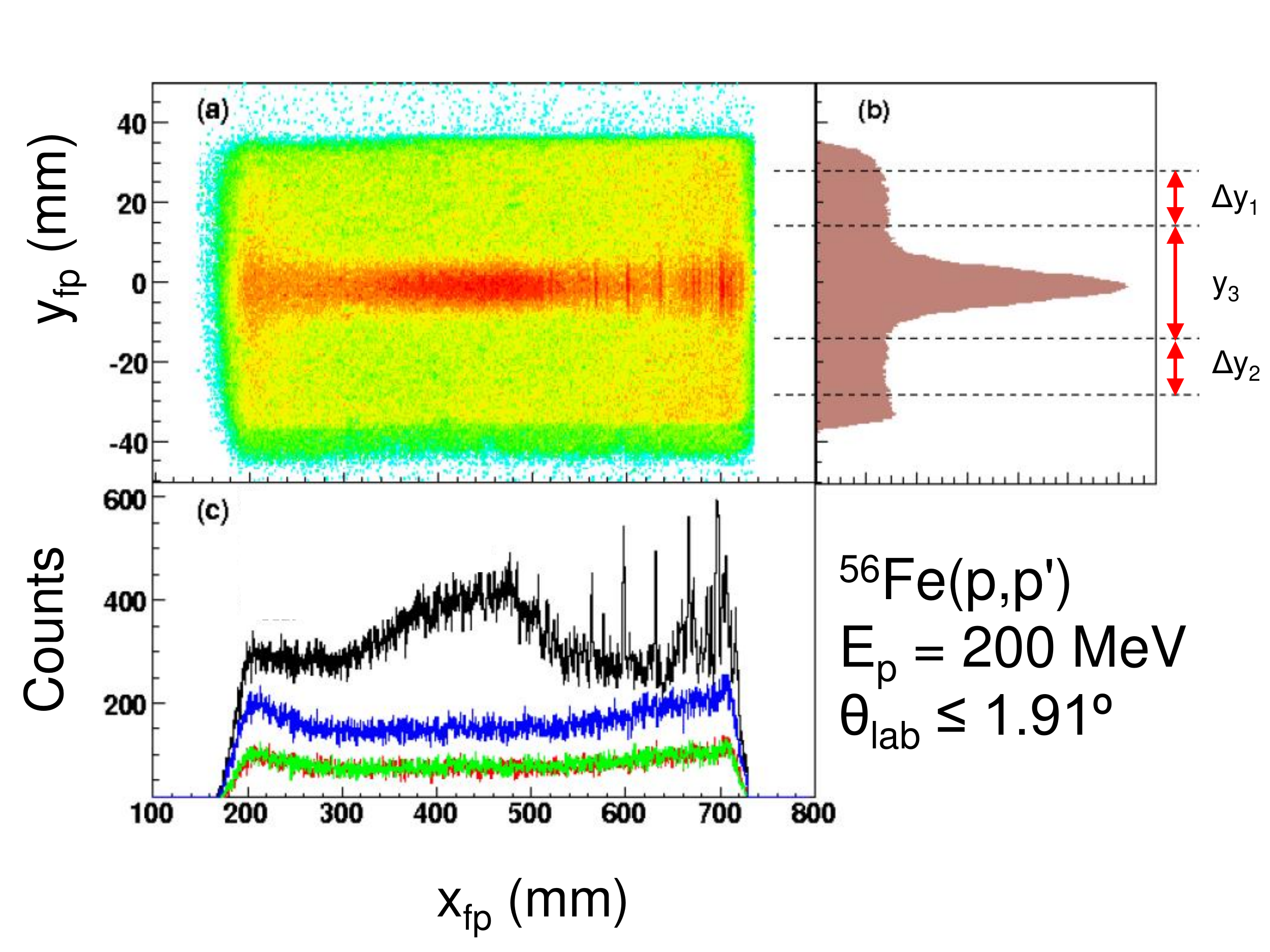}
}
\end{center}
\caption{(a) Two-dimensional scatterplot of focal-plane positions ($x_{\rm fp},y_{\rm fp}$). (b) Projection of the events on the $y_{\rm fp}$ axis with a central gate containing the sum of true and background events and two outer gates representing background events only. The width of the central peak is given by the sum of regions,  $\delta y_{\rm 1}$ and  $\delta y_{\rm 2}$, hence, the total background $y_{\rm 3} = \Delta y_{\rm 1} + \Delta y_{\rm 2}$. (c) Projection on the $x_{\rm fp}$ axis. The black spectrum corresponds to the central gate and the green and red spectra to the background gates. The blue spectrum is the sum of the two background gates. 
\label{fig:xpos2}
}
\end{figure*}
Because of the extreme forward scattering angles of the measurement the dominant source of physical background is small-angle (Coulomb) scattering in the target.
Therefore, the spectrometer was operated in a vertical focus mode, where the protons of interest are focused around the vertical focal plane postion ($y_{\rm fp} = 0$), while the background and remaining beam halo events are expected to be evenly distributed in the vertical direction. 
A two-dimensional scatterplot of horizontal  ($x_{\rm fp}$) and vertical ($y_{\rm fp}$) coordinates in the focal plane for proton inelastic scattering of $^{56}$Fe at $E_{\rm p}$ = 200 MeV is displayed in fig.~\ref{fig:xpos2}(a). 
A projection on the vertical axis (fig.~\ref{fig:xpos2}(b)) shows an enhancement of events around $y_{\rm fp} = 0$ on top of an approximately constant background.
Gates for two background regions above and below the peak at $y_{\rm fp} = 0$ are indicated.
Projection onto the horizontal axis for the central region and two backgrounds regions are shown as black, red and green lines, respectively, in fig.~\ref{fig:xpos2}(c).
The good agreement between the red and green spectra justifies the assumption that the background distribution is independent from the position in the non-dispersive direction.
The blue spectrum in fig.~\ref{fig:xpos2} represents the sum of both background regions.

\begin{figure}[tbh]
\begin{center}
\resizebox{0.45\textwidth}{!}{%
\includegraphics{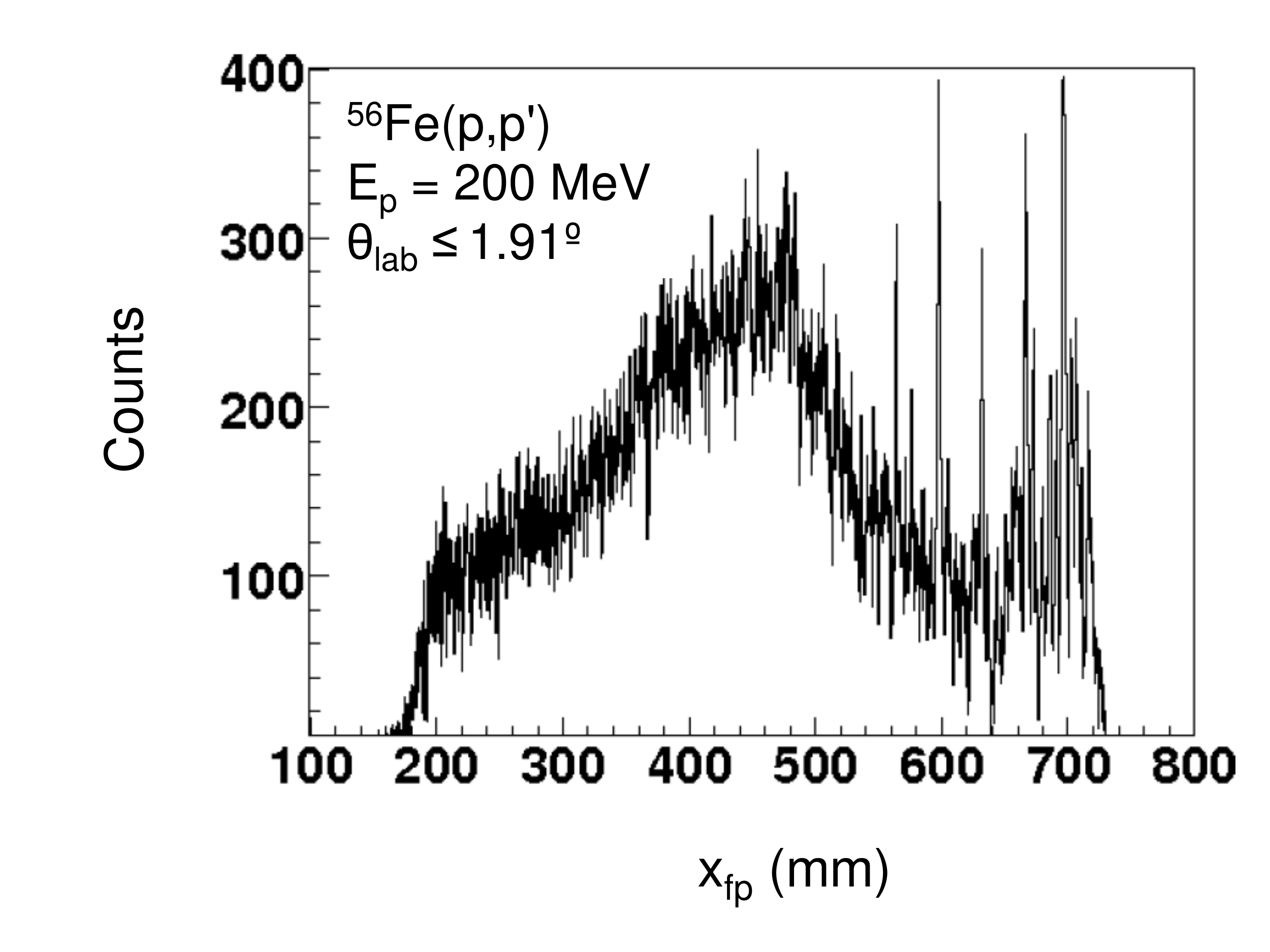}
}
\end{center}
\caption{Background-subtracted spectrum of the $^{56}$Fe(p,p$'$) reaction applying the gates shown in  fig.~\ref{fig:xpos2}.
Note that excitation energy in $^{56}$Fe increases with decreasing $x_{\rm fp}$. 
 \label{fig:xposdiff}
}
\end{figure}

A background-subtracted spectrum is shown in fig.~\ref{fig:xposdiff} for the example of the $^{56}$Fe(p,p$'$) reaction.
Note that the position in the dispersive focal plane increases with the momentum of the scattered particle, i.e. excitation energy increases with decreasing $x_{\rm fp}$. 

\subsection{Lineshape correction}
\label{sec:line}

\begin{figure}[t]
\begin{center}
\resizebox{0.5\textwidth}{!}{%
\includegraphics{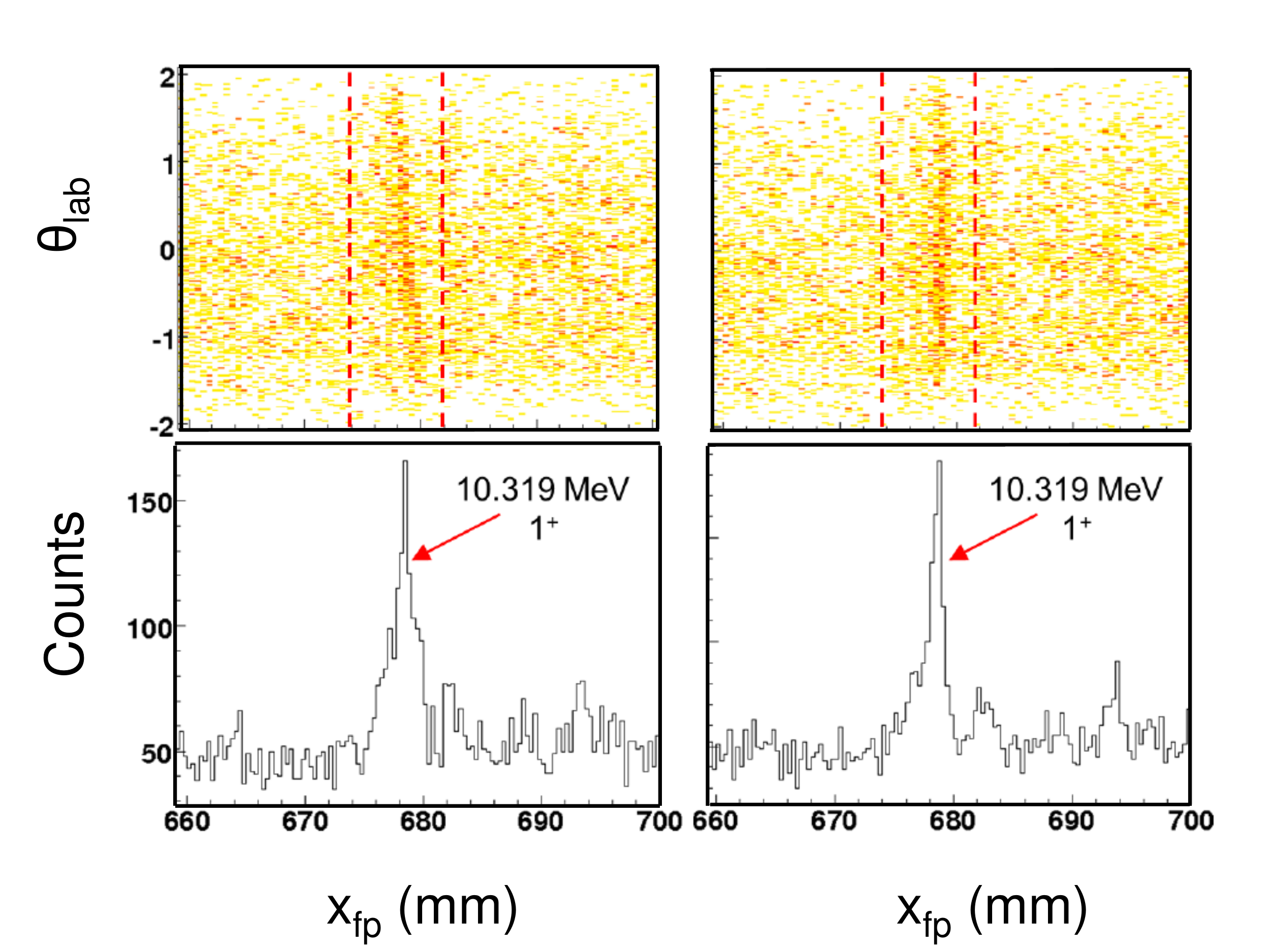}
}
\end{center}
\caption{L.h.s.: Two-dimensional scatterplot of $\theta_{\rm scat}$ \textit{versus} \ $x_{\rm fp}$ around the peak at 10.319 MeV strongly populated in the $^{40}$Ca(p,p$'$) reaction (top) and projection on the dispersive focal plane (bottom).
R.h.s.: Same after application of the lineshape corrections, eqs.~(\ref{eq:ls1},\ref{eq:ls2}).
 \label{fig:corr}}
\end{figure}

Owing to the kinematics of the reaction, particles emerging from the target at the same excitation energy of the residual nucleus but at different reaction angles have different momenta. 
This, together with optical aberrations of the K600 spectrometer, can influence the energy resolution if the angular acceptance is large (as in the present case). 
Experimentally, coils providing higher-order magnetic fields are adjusted in the spectrometer dipole to minimise these effects. 
The left-hand side (l.h.s.) of fig.~\ref{fig:corr} shows the effects of kinematics and aberrations on the energy spectra of the present experiment for the example of the well-known, strong  transition populating the $1^+$ state at 10.319 MeV in $^{40}$Ca \cite{gro79}.
A dependence of the focal-plane position $x_{\rm fp}$ on the scattering angle is clearly visible which leads to a broadening in the energy spectrum (the projection on the  $x_{\rm fp}$ axis) shown in the lower plot leading to a typical resolution of 80 keV (FWHM). 
This effect was removed with a lineshape correction by adding a fourth-order polynomial function of $\theta_{\rm scat}$ to the original $x_{\rm fp}$ value until the curvatures were removed 
\begin{equation}
x_{\rm fpcorr} = x_{\rm fp} - \sum_{n = 0}^4 C_{n} {\theta_{\rm scat}^n}.
\label{eq:ls1}
\end{equation}

The constants $C_n$ were allowed to vary along the dispersive focal plane position ($x_{\rm fp}$) as a polynomial of second order
\begin{equation}
C_{n} = \sum_{i=0}^2 a_i  {x_{\rm fp}}^i.
\label{eq:ls2}
\end{equation}
\noindent
Here, the parameters were determined by a least-squares fit to a set of known transitions in $^{12}$C and $^{27}$Al. For further details see ref. \cite{nev11}.
The right-hand side (r.h.s.) of fig.~\ref{fig:corr} demonstrates the effect of the lineshape correction. 
The energy resolution is considerably improved to 45 keV (FWHM).

\subsection{Resulting spectra}
\label{subsec:spectra}

\begin{figure}
\begin{center}
\resizebox{0.5\textwidth}{!}{%
\includegraphics{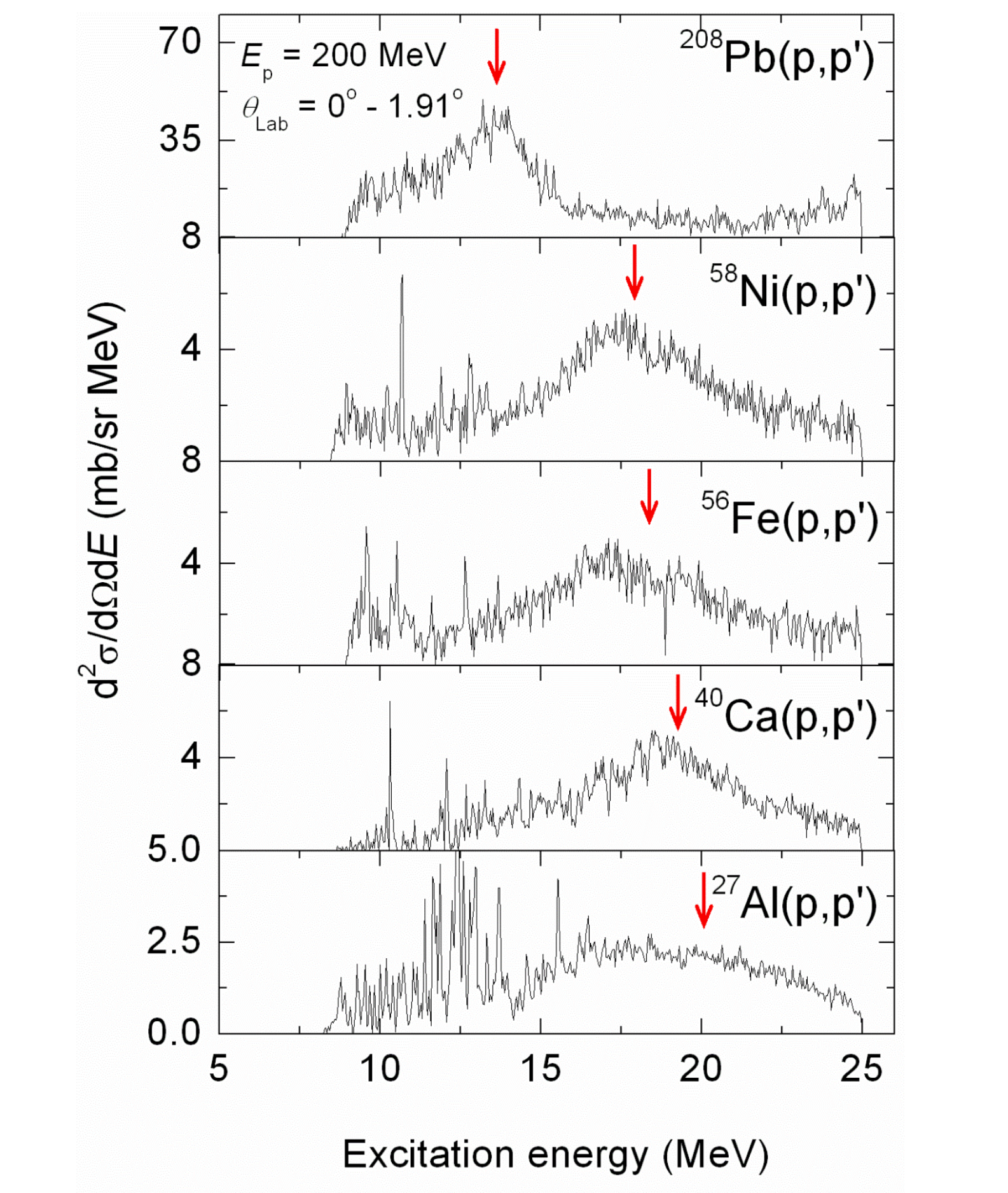}
}
\end{center}
\caption{Experimental double differential cross sections for the (from top to bottom) $^{208}$Pb,$^{58}$Ni,$^{56}$Fe,$^{40}$Ca,$^{27}$Al(p,p$'$) reactions at $E_{\rm p}$ = 200 MeV and scattering angles $\theta_{\rm lab} = 0^{\circ} - 1.91^{\circ}$. The red arrows show locations of the IVGDR across the spectra. \label{Fig:diff}}
\end{figure}

The spectra resulting from the procedures described above were converted to double differential cross sections.
These differential cross sections are displayed in fig.~\ref{Fig:diff} with decreasing mass number from top to bottom.
The maximum of the prominent bump visible in all data follows the systematics of the IVGDR \cite{har01} 
\begin{equation}
E_{\rm C} = 31.2 \, A^{-1/3} + 20.6 \, A^{-1/6},
\label{eq:gdrec}
\end{equation}
except for $^{27}$Al.
However, the IVGDR in light-mass nuclei is known to be extremely fragmented \cite{era86} so that part of the $E$1 strength is likely to be outside the momentum acceptance of the spectrometer. 
Pronounced fine structure is visible over the excitation energy region of the IVGDR in all nuclei investigated, thus confirming the global character of this phenomenon.
The overall cross sections decrease considerably with atomic number indicating that they are dominated by Coulomb excitation. 

\section{Spectrum decomposition and photoabsorption cross sections}

In this section we discuss an attempt to decompose the spectra into the contributions from the Coulomb and nuclear interactions.
A method to convert the double differential cross sections due to Coulomb excitation into equivalent photoabsorption cross sections is presented. 
This allows to determine phenomenologically the nuclear background, whose parameters are adjusted by comparison to photoabsorption data. 

\subsection{Nuclear background}
\label{subsec:backgr}

Because of the selectivity at extreme-forward scattering angles, the background can be approximated by two major sources,  (i) the ISGQR component with a resonance energy given by $E_{\rm x}$ = $63A^{-1/3}$ MeV \cite{har01} placing it underneath the IVGDR and (ii) a component rising towards higher excitation energies. 
The contributions from the ISGQR can be parameterised assuming a modified Lorentzian shape (see e.g.\ ref.~ \cite{car71})
\begin{equation}
\sigma(E) \propto  \frac{\rm \Gamma^2_{\rm R}\textit{E}^2}{(E^2 - E_{\rm R}^ 2)^2 + \rm \Gamma^2_{\rm R}\textit{E}^2}, \label{eq:intro1}
\end{equation}
where $E_{\rm R}$ is the resonance centroid energy and $\rm \Gamma_{\rm R}$ is the resonance half-width. 
The second background component (ii) may contain contributions from quasi-free scattering and the excitation of low angular-momentum modes at higher excitation energies such as the IVGQR or the spin-dipole resonance.
In order to reduce the number of parameters, component (ii) was also assumed to be of the form given by eq.~(\ref{eq:intro1}), and the peak energies were taken from the systematics \cite{har01} of the ISGQR and IVGQR, respectively, while the widths and maximum cross sections were adjusted such that the spectrum converted to equivalent photoabsorption cross sections assuming Coulomb excitation reproduced photoabsorption data. 

It should be noted that in the mass region investigated 27 $\leq A \leq$ 58 the strength distribution of the Isoscalar Giant Monopole Resonance (ISGMR) strongly overlaps that of the ISGQR and only a fraction of $E$0 EWSR is present \cite{har01}. 
In addition, as for the (p,p$'$) experiment, at scattering angles close to $0^{\circ}$ the cross section is dominated by the IVGDR and scaling from DWBA $L$ = 0,2 calculations (cf.~ref.~\cite{don18}) limits the $E$0/$E$2 contribution to less than 10\% of the IVGDR. 
In the case of $^{208}$Pb, the ISGMR moves to the top end of the region of excitation of the IVGDR into component (ii) as an additional multipole contribution. 

\begin{figure}
\begin{center}
\resizebox{0.5\textwidth}{!}{%
\includegraphics{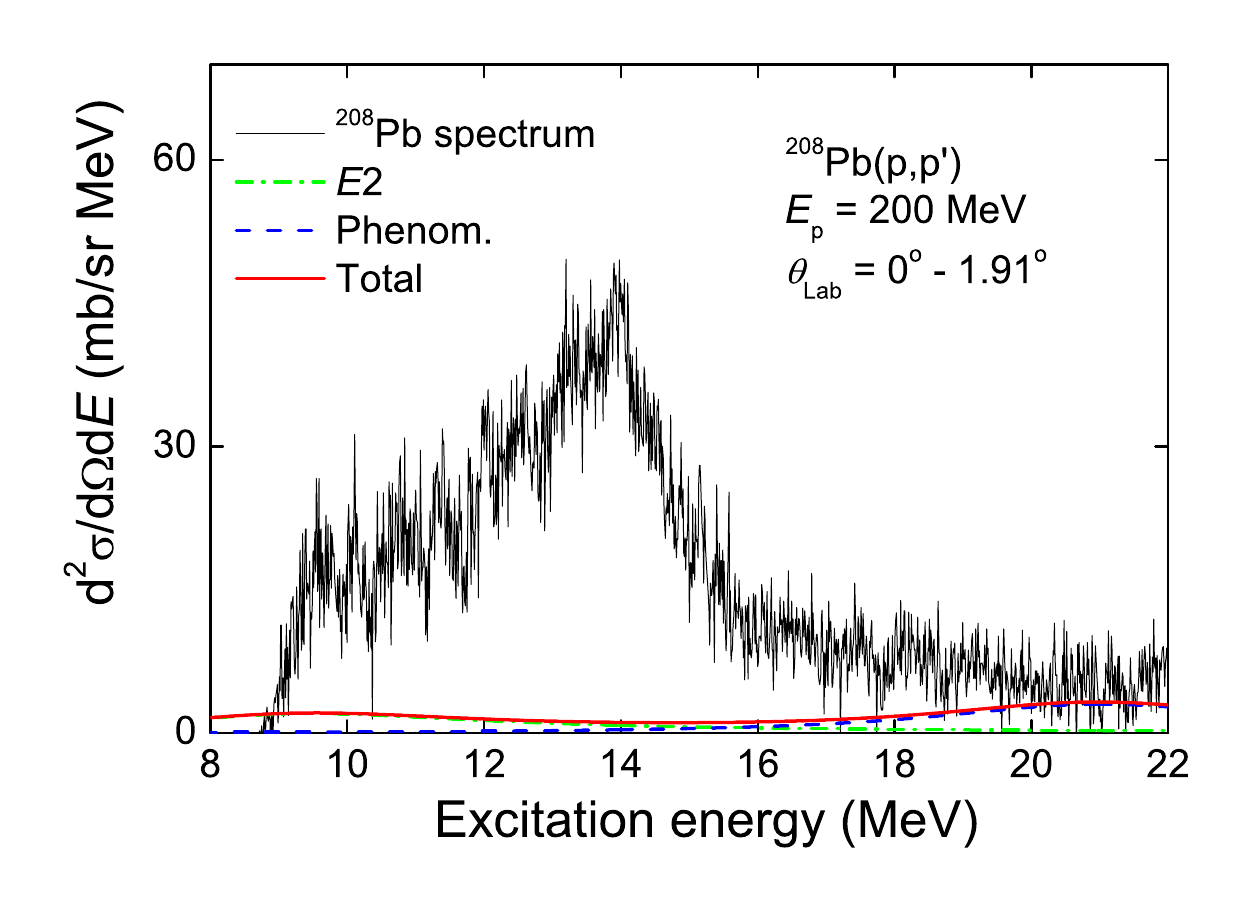}
}
\resizebox{0.5\textwidth}{!}{%
\includegraphics{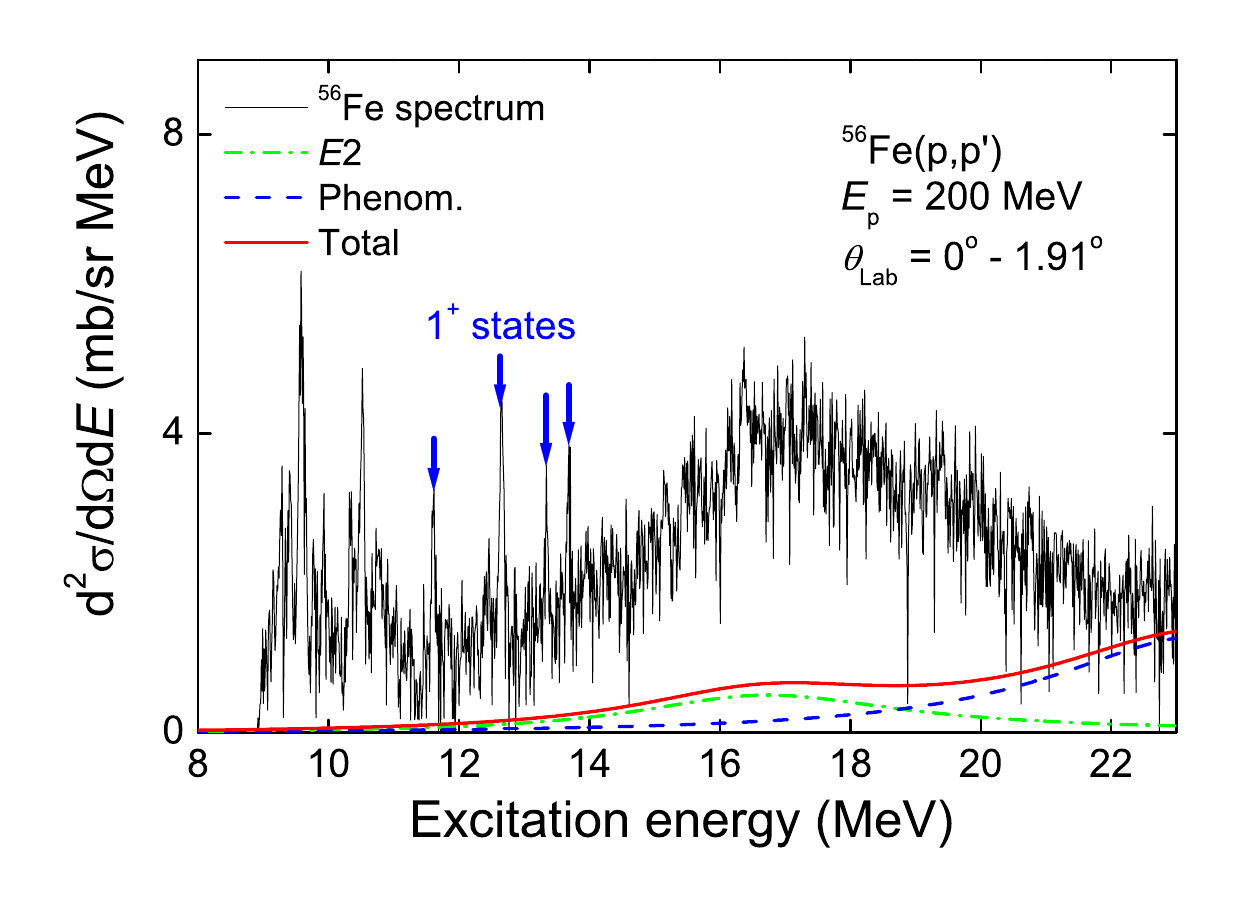}
}
\end{center}
\caption{Top panel: Background contributions to the excitation-energy spectrum of the $^{208}$Pb(p,p$'$) reaction in the IVGDR region by \textit{E}2 cross sections (green dash-dotted curve) and a phenomenological background (blue dashed curve) and their sum (full red curve). 
Bottom panel: Same for $^{56}$Fe. 
Candidates of \textit{M}1 transitions \cite{nag09} are indicated by blue arrows. \label{fig:subt}}  
\end{figure}

Figure~\ref{fig:subt} illustrates the resulting background contributions for the examples of $^{208}$Pb and $^{56}$Fe.
The ISGQR component corresponds to a fraction of the cross section that agrees well with DWBA calculations assuming exhaustion of the energy-weighted sum rule (EWSR). 
For $^{208}$Pb, the present data can be compared with a similar measurement at a higher incident proton energy of 295 MeV \cite{tam11,pol12,pol14}.
There, the phenomenological background could be determined from a multipole decomposition based on angular distributions. 
This phenomenological component also contains a contribution from quasi-free scattering \cite{bir17}.
The resulting background is very similar in shape and magnitude to the one shown in fig.~1 of ref.~\cite{pol14}, in agreement with an expected approximate constancy of the nuclear cross sections as a function of incident proton energy. 
The prominent transitions on the low-energy tail of the IVGDR in $^{56}$Fe are likely of $M1$1 character with isospin $T = T_{\rm g.s.} +1$ \cite{nag09}.
Their long lifetime results from isospin symmetry which inhibits of the usually dominant neutron decay channel.

\subsection{Equivalent photon method}
\label{subsec:phot}

\begin{figure}
\begin{center}
\resizebox{0.5\textwidth}{!}{%
\includegraphics{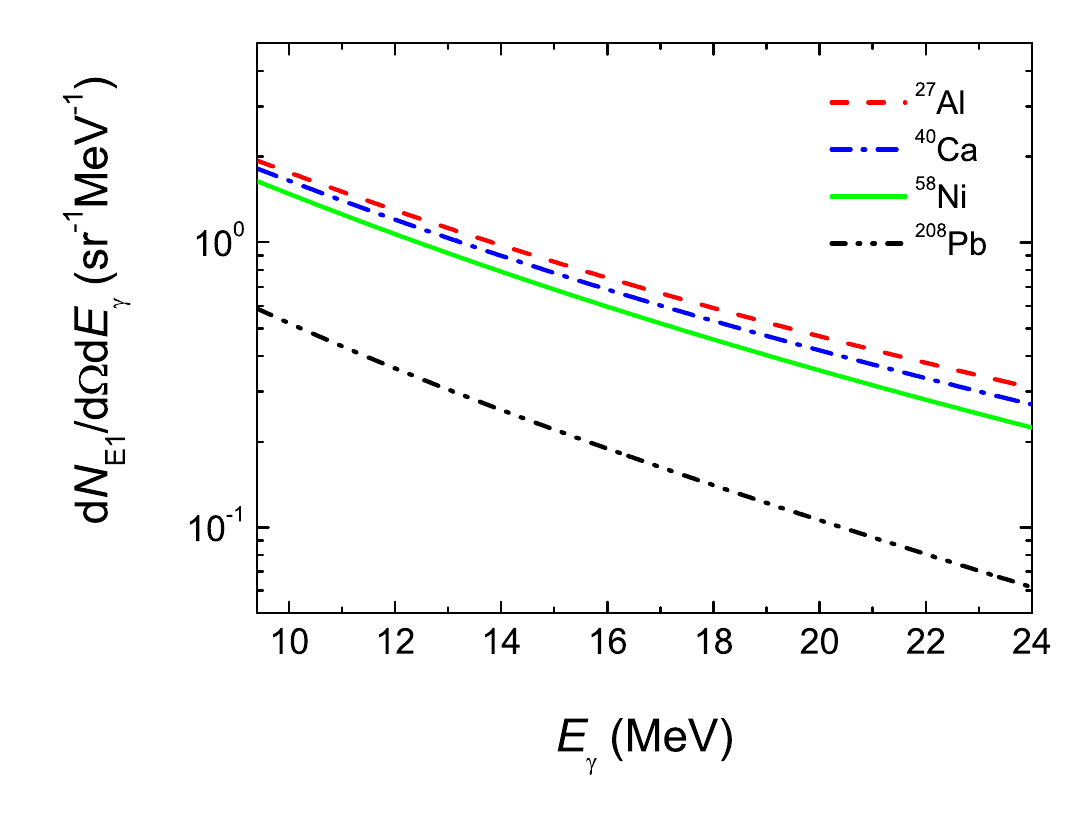}
}
\end{center}
\caption{Total number of virtual photons for the $E$1 multipolarity, as created by a proton passing by the experimental targets at an incident energy of $E_{\rm p}$ = 200 MeV. $^{56}$Fe is not shown as it gives an almost identical result to $^{58}$Ni. \label{fig:produc}} 
\end{figure}

In order to compare the previously published total photoabsorption data with the present measured results, there is a need to convert the measured (p,p$'$) cross sections  into equivalent photoabsorption cross sections \cite{ber88}.
According to the equivalent photon method, the excitation of the target nucleus can be described as the absorption of equivalent photons whose spectrum is determined by the Fourier transform of the time-dependent electromagnetic field generated by the projectile \cite{jac75}. 
The method is used to calculate the equivalent photon numbers for $E$1 multipolarity of the virtual radiation.
The virtual $E$1 photon spectrum \cite{ber88} for each target was calculated using the the Eikonal approximation \cite{ber93} and averaged over the angular acceptance of the detector. 
The equivalent photoabsorption spectrum was obtained using the equation
\begin{equation}
\dfrac{\rm d^{2}\sigma}{\rm d\Omega \rm d\textit{E}_{\gamma}} = \dfrac{1}{E_\gamma}\dfrac{ \rm d\textit{N}_{\textit{E}1}}{\rm d\Omega\rm d\textit{E}_{\gamma}}\sigma{_{\gamma}^{\pi \lambda}}(E_{\gamma}). \label{eq:virt1}
\end{equation}
The resulting \textit{E}1 virtual photon numbers for the various targets are shown in fig.~\ref{fig:produc}. 
\subsection{Comparison to total photoabsorption measurements}
\label{subsec:electro}

\begin{figure}
\begin{center}
\resizebox{0.5\textwidth}{!}{%
\includegraphics{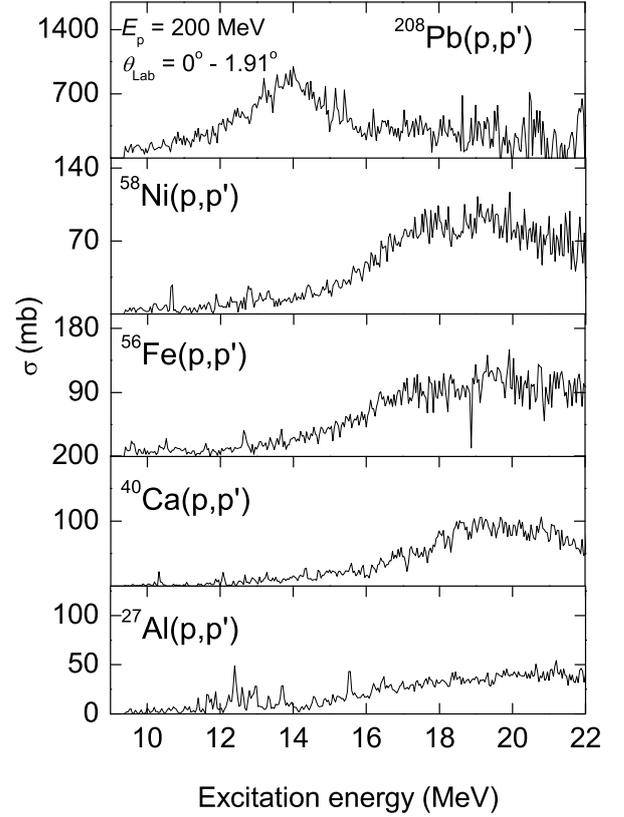}
}
\end{center}
\caption{Coulomb excitation cross sections in the (top to bottom) $^{208}$Pb, $^{58}$Ni, $^{56}$Fe, $^{40}$Ca, $^{27}$Al(p,p$'$) reactions at $E_{\rm p}$ = 200 MeV and $\theta_{\rm lab} = 0^{\circ} - 1.91^{\circ}$. \label{fig:Ca-q}}
\label{fig:5}       
\end{figure}

After the subtraction of the nuclear background components described in sect.~\ref{subsec:backgr}, all spectra were converted from Coulomb excitation to photabsorption cross sections with the procedure discussed above (see fig.~\ref{fig:Ca-q}). 
These spectra equivalent to the \textit{E}1 response spectra better represent the IVGDR in terms of position and width than the spectra of fig.~\ref{Fig:diff}.\\

In order to obtain equivalent photoabsorption spectra, one needs to multiply the equivalent $E$1 response in each energy bin with the energy 
\begin{equation}
\sigma_{i\rightarrow f} =\sum_{\pi \lambda}\int n_{\pi \lambda}(\textit{E}_{\gamma})\sigma_{\gamma}^{\pi \lambda}(\textit{E}_{\gamma})\frac{d\textit{E}_{\gamma}}{\textit{E}_{\gamma}},
\end{equation}
where the spectrum of photons of multipolarity $\lambda$ is determined by the equivalent photon number $n_{\pi \lambda}$ and the photoabsorption cross section is given by $\sigma_{\gamma}^{\pi \lambda}$.\\ 

\begin{figure}
\begin{center}
\resizebox{0.5\textwidth}{!}{%
\includegraphics{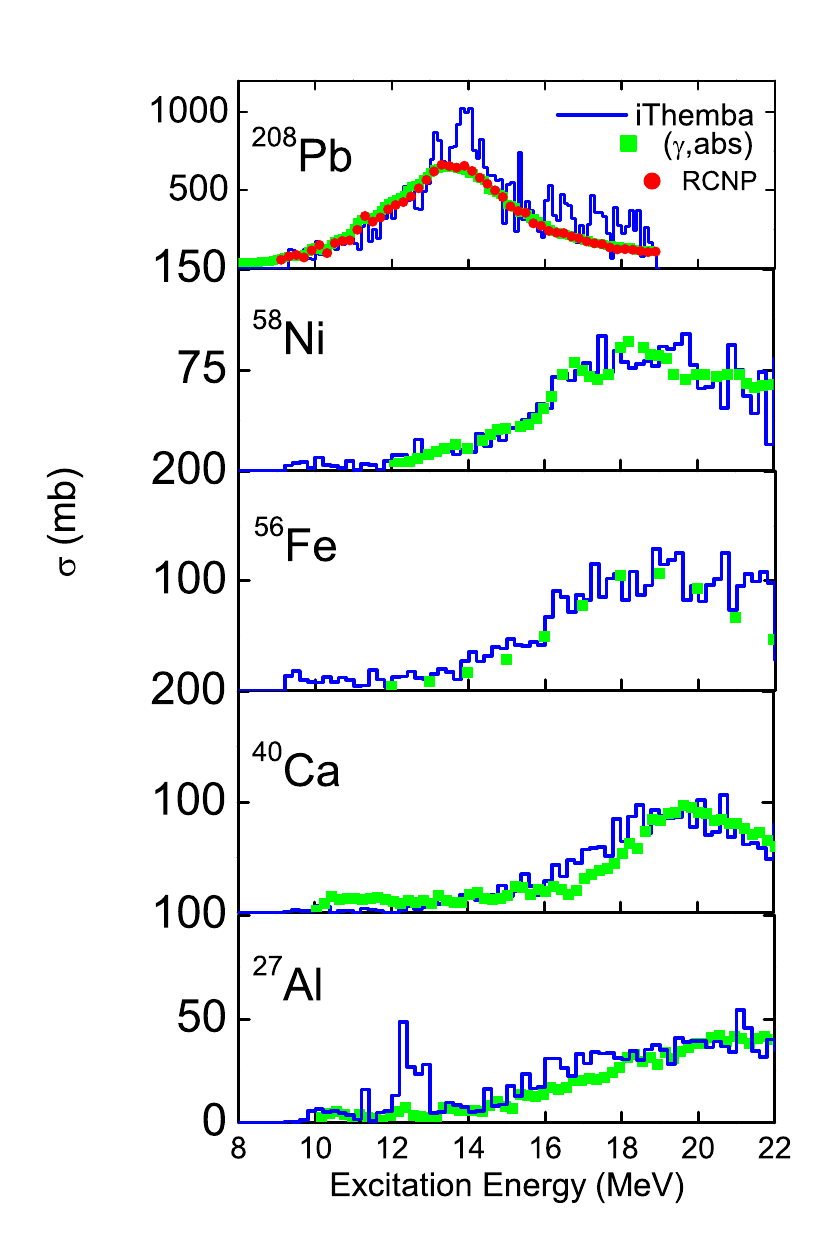}
}
\end{center}
\caption{Top to Bottom panels: Equivalent photoabsorption cross sections obtained in the present work at $E_{\rm p}$ = 200 MeV for $^{208}$Pb (blue histogram) in comparison to  $\sigma$($\gamma$,abs) data from \cite{vey70,sch88} (green squares) and RCNP work at $E_{\rm p}$ = 295 MeV \cite{tam11} (red circles). 
Same for $^{58}$Ni in comparison to ref.~\cite{ish02}, $^{56}$Fe in comparison to ref.~\cite{bor00}, $^{40}$Ca  and $^{27}$Al in comparison to ref.~\cite{ahr75} (green squares).
\label{fig:dd}}
\end{figure}

Figure~\ref{fig:dd} displays a comparison between the converted (p,p$'$) equivalent photoabsorption data and photoabsorption measurements for $^{208}$Pb \cite{tam11,vey70,sch88}, $^{58}$Ni \cite{ish02}, $^{56}$Fe \cite{bor00}, $^{40}$Ca \cite{ahr75} and $^{27}$Al \cite{ahr75}. 
The (p,p$'$) equivalent photoabsorption results of the nuclei investigated were found to be consistent with previously published data. 
However, absolute photoabsorption cross sections from the present experiment have large uncertainties. 
The present setup at $\theta_{\rm lab} = 0^{\circ}$ does not allow for the determination of accurate vertical scattering angles, thus limiting the angular resolution (see also refs.~\cite{nev11,don18}). 
Consistency between the two data sets was achieved by scaling the equivalent photoabsorption cross sections to the published data. 
Normalisation factors $ 0.6 \pm 0.1$ were used for all targets, similar to within uncertainty to those reported in ref.~\cite{don19} for Nd and Sm isotopes using the same setup.

While the photoabsorption measurements agree well amongst each other (see also ref.~\cite{tam11}), the present experiment finds somewhat smaller values at lower excitation energies for $^{208}$Pb. 
However, the maximum and the higher-energy tail of the IVGDR are well reproduced quantatively.
It should be noted that the $^{208}$Pb data set has limited statistics and due to its high charge $Z = 82$ the experimental background from small-angle atomic scattering in the target was much more severe than for the other targets.
The $^{58}$Ni($\gamma$,abs) cross sections \cite{ish02} are in remarkable agreement with results deduced from the present data suggesting that the Eikonal approximation calculation of eq.~(\ref{eq:virt1}) is indeed able to reproduce previously published photoabsorption data. 
There is also acceptable agreement for the cases of $^{40}$Ca and $^{27}$Al, where the results from the virtual photon method can be compared to the data from ref.~\cite{ahr75}.

\section{Fine structure of the IVGDR and characteristic scales}

\subsection{Wavelet analysis}

\begin{figure*}
\begin{center}
\resizebox{0.9\textwidth}{!}{%
\includegraphics{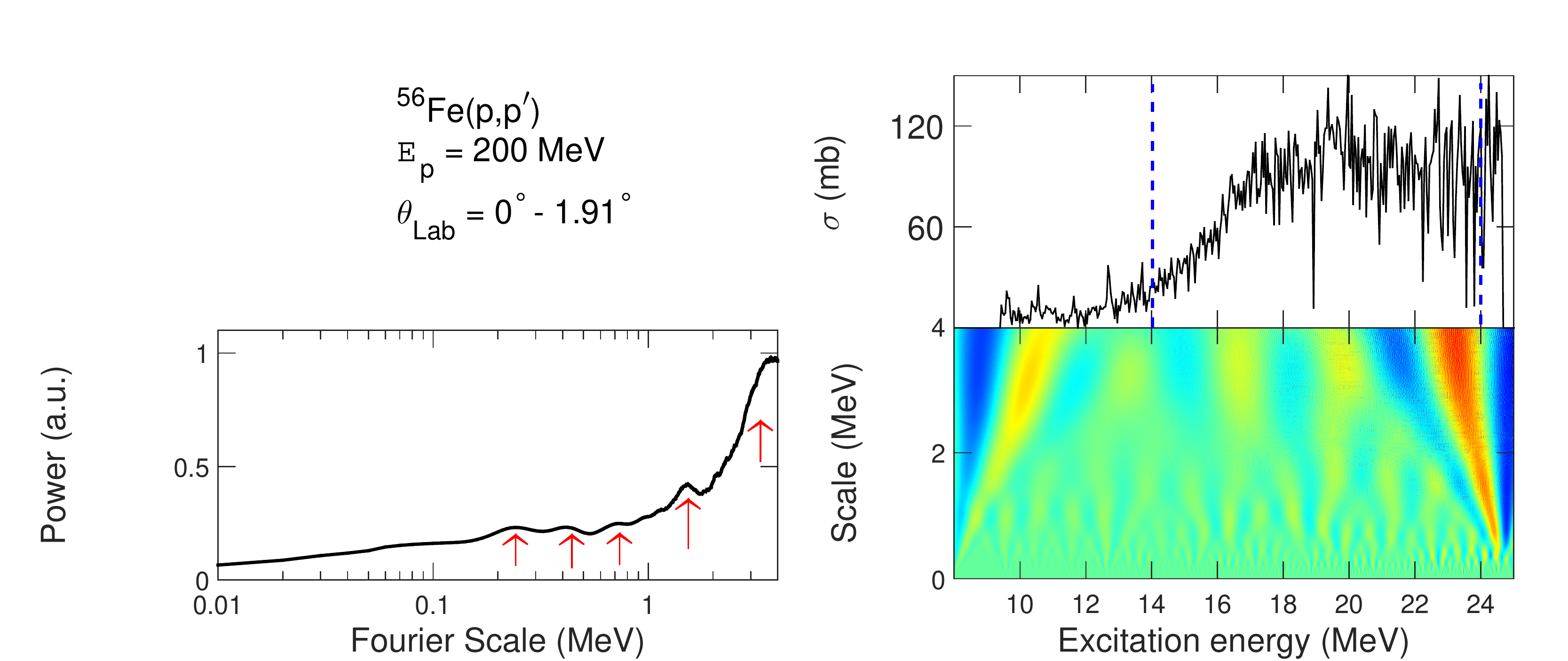}
}
\end{center}
\caption{R.h.s.: CWT analysis of the equivalent photoabsorption spectrum of the $^{56}$Fe nucleus. 
 L.h.s.: The power spectrum representing the region 14 $\leq E_{\rm x} \leq$ 24 MeV. The CWT plot (R.h.s.) displays wavelet scale which is equivalent to the Fourier scale (l.h.s.). \label{fig:Fe-exp}}
\end{figure*}

As discussed in the introduction, the wavelet analysis technique has been used extensively for the extraction of characteristic energy scales from the fine structure of giant resonances. 
The wavelet transform is a flexible tool particularly useful for the analysis of transients, aperiodicity and other non-stationary signal features. It is classified into two types: the Continuous Wavelet Transform (CWT) and the Discrete Wavelet Transform (DWT).  
The two wavelet transforms have one common use, they can be utilised in the representation of continuous position spectra. 
In general, the CWT can be applied over every possible scale and translation while on the other hand, the DWT is applicable only on a specific subset of scale and translation values or representation grid.

In the present case, energy spectra of nuclear giant resonances are analysed.
The coefficients of the wavelet transform are defined as \cite{matl}
\begin{equation}
   C\left( {\delta E,E_{\rm x}} \right) = \frac{1}{\sqrt{\delta E}}\int {\sigma \left( E \right)\Psi^{*} \left( \frac{E_{\rm x} - E}{\delta E}
   \right)dE},
   \label{eq:cwt}
\end{equation}
where the coefficients depend on two parameters, the scale $\delta E$ stretching and compressing the wavelet $\Psi (E)$, and the position $E_{\rm x}$ shifting the wavelet in the spectrum $\sigma (E)$. 
The analysis of the fine structure of giant resonances is performed using the CWT, where the fit procedure can be adjusted to the required precision. 
Applications of the CWT to high energy-resolution nuclear spectra of giant resonances are described in refs.~\cite{she04,she09,pol14,usm11,pet10,usm16,fea18}. 
Further details and a comparison with other techniques for the analysis of fine structure in nuclear giant resonances can be found in ref.~\cite{she08}.

In order to achieve an optimum representation of the signal using wavelet transformation, one has to select a wavelet function $\Psi$ which resembles the properties of the studied signal $\sigma$. 
A maximum of the wavelet coefficients at  a certain value of $\delta E$ indicates a correlation in the signal at the given scale,  called characteristic scale.
The best resolution for nuclear spectra is obtained with the so-called Morlet wavelet (cf.\ fig.~9 in ref.~\cite{she08}) because the detector response is typically close to the Gaussian line shape and the Morlet wavelet is a product of Gaussian and cosine functions. 
It takes the form
\begin{equation}
\Psi (x) = \frac{1}{\pi^{\frac{1}{4}}}\rm cos(\textit{ikx})\rm exp\left(-\frac{\textit{x}^{2}}{2}\right), \label{wave11}                                                                         
\end{equation}
being evaluated for each scale $\delta \rm E$ and position $E_{\rm x}$, with the parameter \textit{k} weighing the resolution in scales \textit{versus} the resolution in localisation. In order to satisfy the admissibility conditions $k \lessapprox 5$ must be fulfilled \cite{far92}. 
In this present analysis the complex Morlet Mother wavelet \cite{teo98} was used. 

\subsection{Application to the (p,p$'$) spectra}

The application of the CWT to the present experimental results in the excitation energy region of the IVGDR is illustrated for the example of $^{56}$Fe in fig.~\ref{fig:Fe-exp}. 
The top-right panel represents the equivalent photoabsorption spectrum and bottom-right panel displays the corresponding wavelet coefficients  for an energy-scale region up to 4 MeV. 
The colour indicates the magnitude of the wavelet coefficients for a given excitation energy and scale, with blue showing regions of smallest values of the wavelet coefficients, while red represents maximum values. 
Power spectra, for generally better visibility of energy scales, were obtained by taking the square root (at each value of scale) of the usual sum of the squared modulus of the complex CWT coefficients over the excitation energy range of the IVGDR indicated by the vertical dashed blue-lines (see the bottom left-hand side of fig.~\ref{fig:Fe-exp}). 
Clearly visible characteristic energy scales are indicated by red arrows.
\begin{figure*}
\begin{center}
\resizebox{0.9\textwidth}{!}{%
\includegraphics{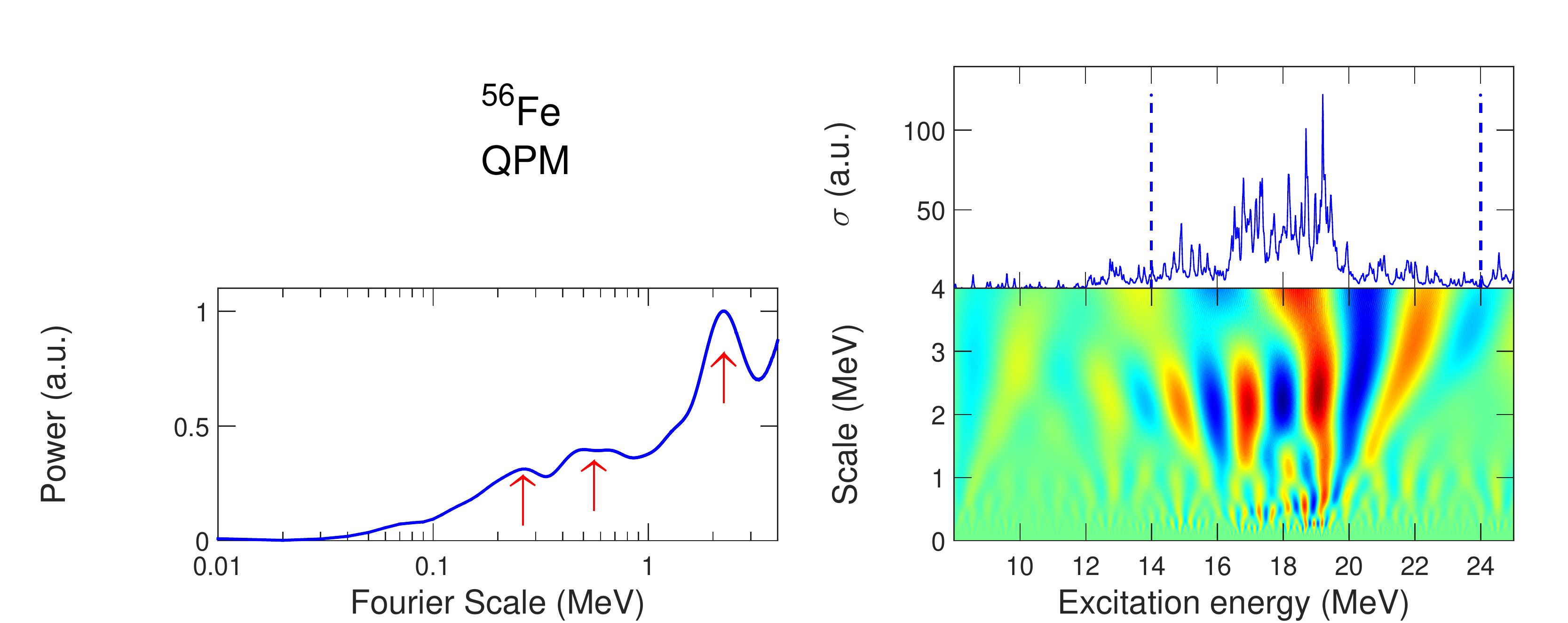}
}
\end{center}
\caption{Same as fig.~\ref{fig:Fe-exp}, but for the QPM calulation of the IVGDR in $^{56}$Fe described in the text.
\label{fig:Fe-qpm}}       
\end{figure*}
 
In order to interpret these energy scales, the equivalent photoabsorption cross sections are compared to QPM calculations.
The QPM has been extremely successful in the description of collective modes at low and high excitation energies  (see e.g.\ refs.~\cite{pon00,bur07,sav08,wal11}) including the IVGDR \cite{rye02}.
Calculations for the IVGDR have been performed on the basis of interacting one- and two-phonon configurations. 
The two-phonon configurations were made up of phonons of spin-parity values $J^\pi =1^{\pm} - 9^{\pm}$. 
They were cut at about 4 -- 5 MeV above the centroid energy of theIVGDR. 
The single-particle basis in QPM calculations is rather complete and includes all mean-field levels from 1$s_{1/2}$ to quasi-bound levels in the continuum. 
For this reason, no effective charges are needed as e.g.\ in shell-model calculations. 
However, a correction is necessary to remove the center-of-mass-motion leading to effective charges $e_{N} = -Z/A$ and $e_{Z} = N/A$. 

Results of the wavelet analysis on QPM $E$1 strength distributions for the $^{56}$Fe nucleus are presented in fig.~\ref{fig:Fe-qpm}.
Indeed, the coupling to two-phonon states leads to a strong fragmentation, as demonstrated below, which resembles the width of the resonance. 
The power spectrum again shows characteristic maxima to be compared with experiment.
They also display a continuous increase with scale value. 
However, this is a trivial effect due to the large overlap of the wavelet function with the broad resonance structure in the integral of eq.~(\ref{eq:cwt}).

\subsection{Characteristic energy scales of $^{40}$Ca, $^{56}$Fe, $^{58}$Ni and $^{208}$Pb}
\begin{figure*}
\begin{center}
\resizebox{0.75\textwidth}{!}{%
\includegraphics{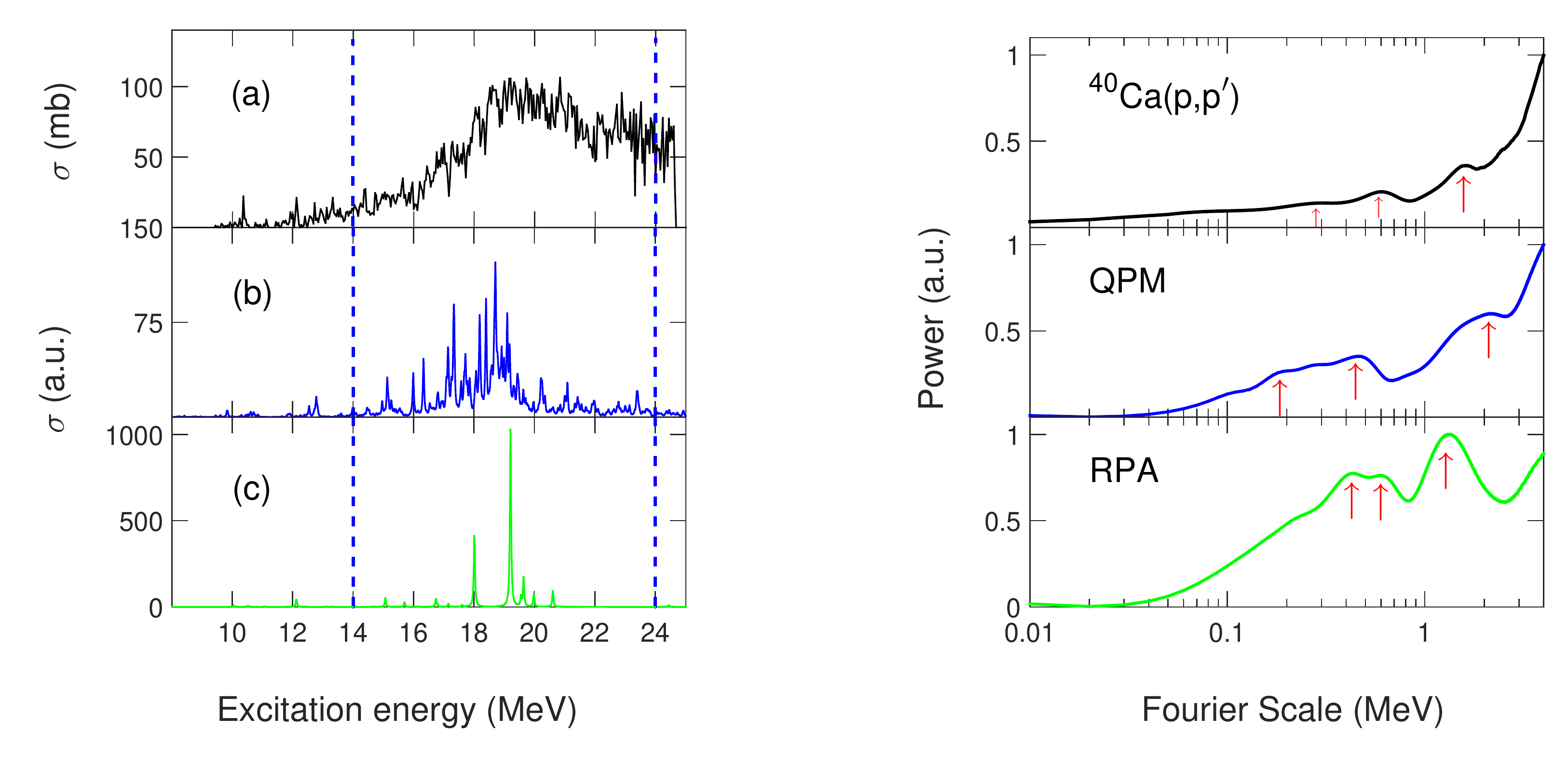}
}
\end{center}
\caption{Left: Equivalent photoabsorption spectrum of $^{40}$Ca (a) in comparison with QPM predictions in (1+2)-phonon (b) and 1-phonon (c) model spaces. Right: Power spectra from the  CWT analysis integrated over the energy region 14 $\leq E_{\rm x} \leq$ 24 MeV.
\label{fig:Ca-rpa}} 
\end{figure*}
\begin{figure*}
\begin{center}
\resizebox{0.75\textwidth}{!}{%
\includegraphics{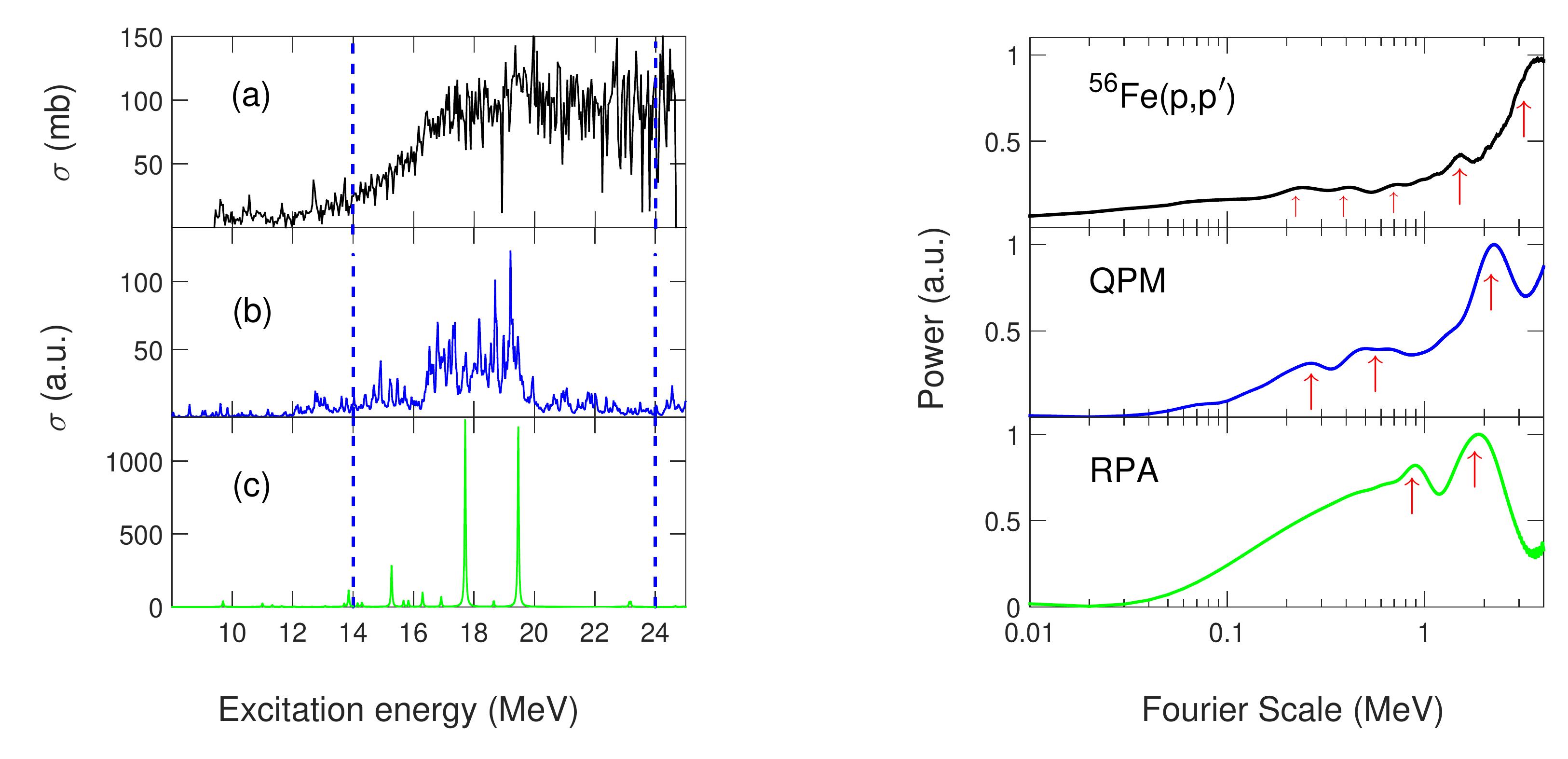}
}
\end{center}
\caption{Same as fig.~\ref{fig:Ca-rpa}, but for $^{56}$Fe. 
\label{fig:Fe-rpa}}       
\end{figure*}
\begin{figure*}
\begin{center}
\resizebox{0.75\textwidth}{!}{%
\includegraphics{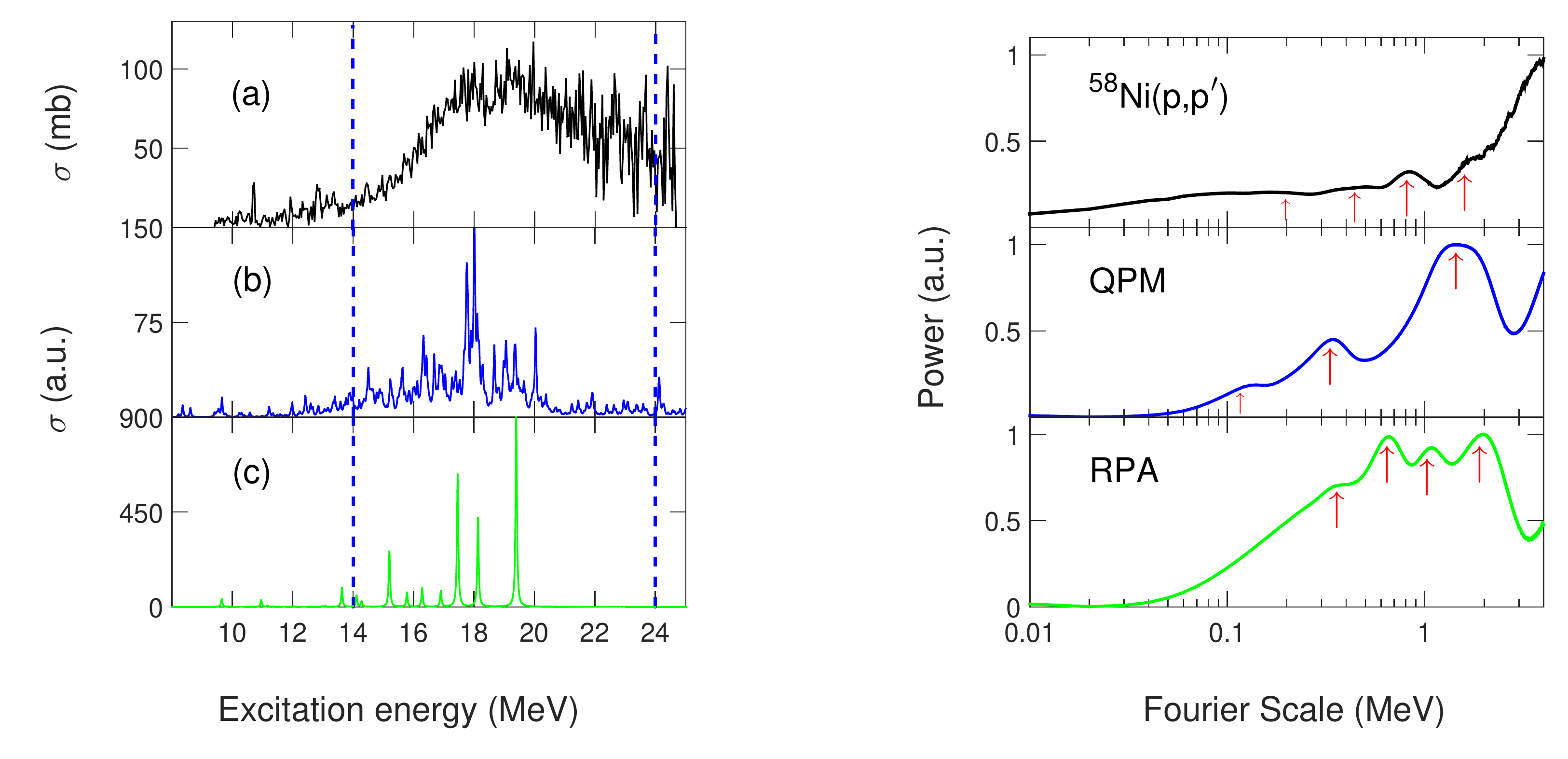}
}
\end{center}
\caption{Same as fig.~\ref{fig:Ca-rpa}, but for $^{58}$Ni.
\label{fig:Ni-rpa}}
\end{figure*}
Figures~\ref{fig:Ca-rpa} -- \ref{fig:Ni-rpa} summarise the results of the  CWT analysis for $^{40}$Ca, $^{56}$Fe and $^{58}$Ni.
Their left-hand sides show the experimental and theoretical photoabsorption cross sections and their right-hand sides the corresponding power spectra.
Two types of calculations are presented, viz.\ on the level of 1-phonon configurations (labelled RPA) and (1+2)-phonon configurations as described above (labelled QPM).  
Clearly,  fragmentation of theIVGDR is already present at the RPA level.
Typically, a distribution into 2 or 3 major and a number of smaller fragments is observed for these $fp$-shell nuclei.  

The power spectra are summed over the excitation energy region 14 $\leq E_{\rm x} \leq$ 24 MeV covering the IVGDR.
The theoretical results display increasing power up to scales of 4 MeV in correspondence with the experimental results but fail to reproduce the further increase to even higher scales. 
This is due to the limitations of the model, which presently does not include all mechanisms contributing to the total decay width.
In particular, direct continuum decay and contributions due to complex states beyond the 2-phonon level are missing.   
   
The spacings between the main fragments lead to pronounced power maxima in the RPA results at scales between about 0.5 MeV and 2 MeV. 
Inclusion of 2-phonon states considerably modifies the power spectra.
Most notably, characteristic scales appear at smaller values between about 100 and 300 keV.
At larger scale values the number of maxima are reduced compared to RPA underlining that a realistic description of fine structure has to go beyond the 1p-1h level.  
\begin{figure*}
\begin{center}
\resizebox{0.75\textwidth}{!}{%
\includegraphics{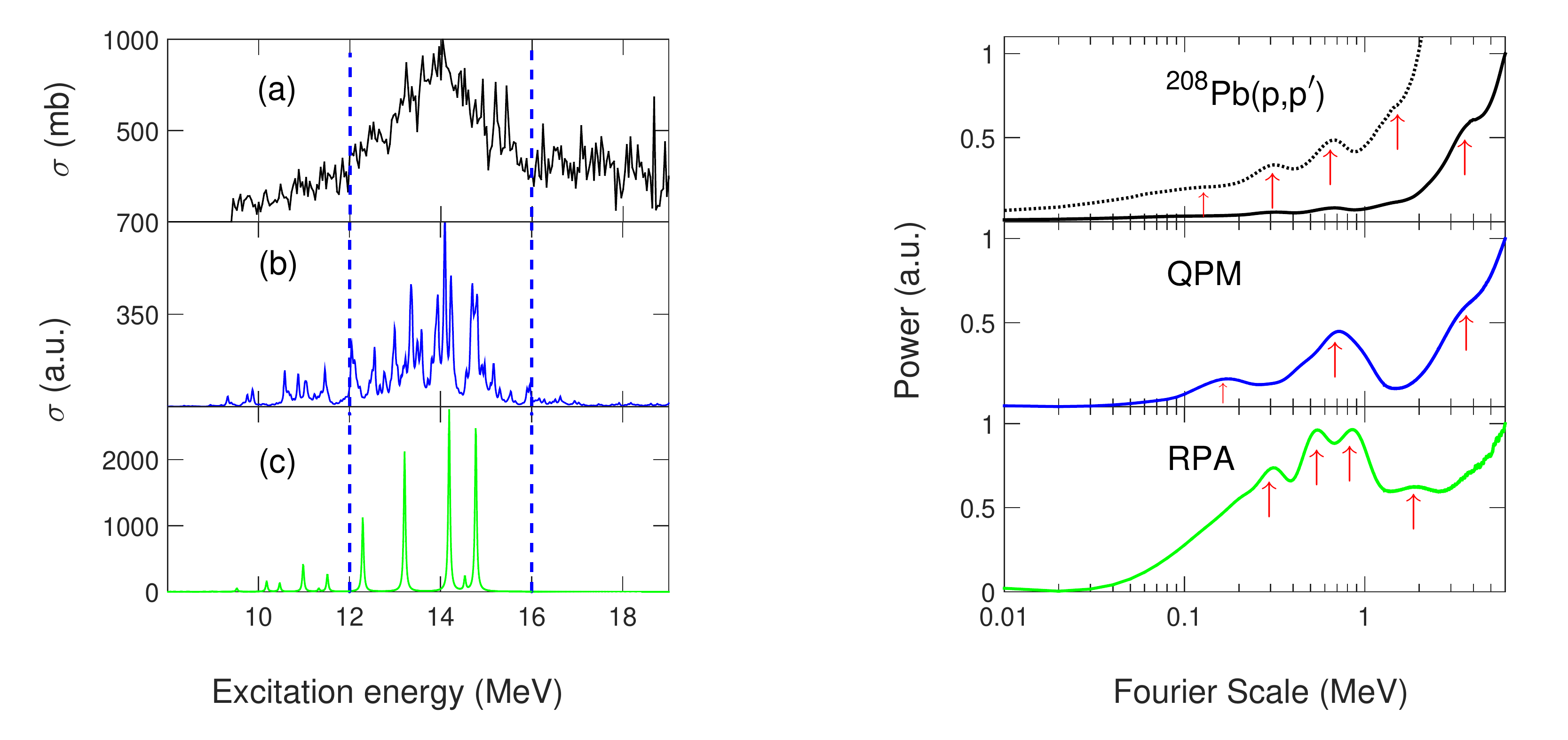}
}
\end{center}
\caption{Same as fig.~\ref{fig:Ca-rpa}, but for $^{208}$Pb and power spectra from the CWT analysis integrated over the energy region 12 $\leq E_{\rm x} \leq$ 16 MeV.
\label{fig:Pb-rpa}}
\end{figure*}

Table \ref{tab:3} summarises the characteristic scales from experiment and calculations indicated by arrows in figs.~\ref{fig:Ca-rpa} -- \ref{fig:Ni-rpa}. 
These are grouped in different Classes with scales $<$ 300 keV for Class I, $300-1000$ keV for Class II and values $>$ 1 MeV for Class III. 
The latter are large scales associated with the overall spread of the strength distributions while Class II are related to both fragmentation on the 1p-1h level and coupling to more complex states.
Class I states can be directly related to the coupling to 2-phonon states.
We note that the choice 300 keV and 2 MeV as borders to distinguish between the different classes is somewhat arbitrary.
The values were chosen to facilitate easy comparison  to previous studies of the IVGDR in $^{208}$Pb \cite{pol14} and of the ISGQR in many nuclei \cite{she04,she09}. 

It should be noted that in a previous work \cite{she08} this group examined in great detail the extraction of scales in CWT analysis. The present use of the Complex Morlet Mother Wavelet leads directly to the equivalent Fourier scale, the well-known "lengthlike" scale (use of the Morlet Mother Wavelet requires division of scale by 0.83 to obtain the equivalent Fourier scale). Conversion of a "widthlike" scale, equivalent to the width of a Lorentzian, used in some of the previous work can be achieved by multiplying that scale by a factor of two to obtain the Fourier scale. As such for uniformity, the values given in table \ref{tab:3} are Fourier scales.  
\begin{table}
\centering
\caption{Characteristic energy scales extracted with the CWT analysis from the investigated nuclei.}
\label{tab:3}       
\begin{tabular}{c c c c c c c c}
\hline\hline\noalign{\smallskip}
&Class I&\multicolumn{4}{c}{Class II}&\multicolumn{2}{c}{Class III}\\
&$<$ 300 keV&\multicolumn{4}{c}{300 -- 1000 keV}&\multicolumn{2}{c}{$>$ 1000 keV}\\
\noalign{\smallskip}\hline\noalign{\smallskip}
\textbf{$^{40}$Ca}\\
Exp.&--&280&--&600&--&--&2000\\
QPM&--&190&440&--&--&--&2000\\
RPA&--&--&440&610&--&1300&--\\
\textbf{$^{56}$Fe}\\
Exp.&--&250&400&640&--&1500&3200\\
QPM&--&250&--&600&--&--&2300\\
RPA&--&--&--&--&840&1800&--\\
\textbf{$^{58}$Ni}\\
Exp.&--&200&--&460&850&1600&--\\
QPM&--&120&350&--&--&1600&--\\
RPA&--&--&350&--&660&1100&2000\\
\textbf{$^{208}$Pb}\\
Exp. &--&130&310&--&670&--&3600\\
\cite{pol14}&160&260&440&860&1280&1920&3500\\
QPM&--&170&--&--&710&--&3600\\
RPA&--&--&310&530&860&1900&--\\
\noalign{\smallskip}\hline
\end{tabular}
\end{table}

The appearance of Class I scales can be directly related to the spreading width.
Although we have no direct proof, the variation of scale values for the different nuclides indicates that coupling to low-lying collective vibrations \cite{ber83}, is the relevant mechanism rather than an average coupling strength between 1p-1h and 2p-2h states.
This mechanism has been identified as major source of scales in case of the ISGQR in medium- to heavy-mass nuclei \cite{she04,she09}.   
For Class II scales the variation of scale values is even more pronounced.
This might be connected to the coupling to low-lying surface vibrations, whose properties change from nucleus to nucleus.
On the other hand, the comparison of RPA and QPM results demonstrates that Landau fragmentation is present and influences the distribution of strength.
The situation is similar to studies of fine structure of the GQR in lighter-mass nuclei \cite{usm11,usm16}, whose interpretation required the same type of interplay.
The reproduction of larger scales (Class III) by the QPM is quite good for $^{40}$Ca and $^{58}$Ni.
In $^{56}$Fe, two experimental scales above 1 MeV are found, while the QPM predicts a single scale only.      
One possible explanation for this difference may be the fact that $^{56}$Fe is considerably deformed, while the QPM model assumes a spherical structure.
Finally, we note that the largest scales corresponding to the total width of theIVGDR could not be extracted with the present approach because of the limited energy window, since the minimum requirement of the CWT is at least two times the value of the scale to identify.

Finally, we briefly discuss the case of  $^{208}$Pb (fig.~\ref{fig:Pb-rpa}).
An experiment similar to the present one, but at a higher proton energy of 295 MeV, has been performed with the Zero-degree Facility at RCNP \cite{tam09} and a detailed analysis of the fine structure of the IVGDR in $^{208}$Pb has been reported in ref.~\cite{pol14}. 
One should be aware that in the present proof-of-principle test of the Zero-degree Facility at iThemba LABS \cite{nev11} the statistics in the spectrum of the heaviest-mass nucleus studied ($^{208}$Pb) was rather poor with corresponding limitations of a CWT analysis.
Nevertheless, acceptable agreement with the results of ref.~\cite{pol14} is achieved if one assumes that the power maxima around 300 and 700 keV each represent in fact two unresolved scales.
At small scales a shallow power maximum is found, which is not as pronounced as in the data of ref.~\cite{pol14}. 
This is most likely due to the limited statistics and the slightly better energy-resolution of the latter data.
The largest scale deduced from both experiments  is in good agreement.
It represents the total width of the resonance, which can be assessed for $^{208}$Pb in contrast to the lighter-mass nuclei discussed above because of the much lower centroid energy of the IVGDR.       

\section{Concluding remarks}
\label{sec:results}

The present work reports results of a proof-of-principle study of the (p,p$'$) reaction at the Zero-degree Facility of the K600 spectrometer \cite{nev11} at iThemba LABS covering a wide mass range of nuclei.
At an incident energy of 200 MeV the cross sections are dominated by relativistic Coulomb excitation thus permitting a detailed study of the IVGDR.
After introduction of empirical procedures to subtract the main contributions from nuclear scattering (excitation of isoscalar giant resonances, quasi-free scattering) one can extract the photoabsorption cross sections.
Although we are not able at present to extract absolute values, the shapes of the deduced photoabsorption spectra generally agree well with results using different techniques.
This is of particular importance in the light of the surprising results of a subsequent study of the stable Nd isotopes representing a transition from spherical nuclei to prolate deformed rotors \cite{don18}.
These experiments, using the same setup, showed large differences in the energy distribution compared with a previous ($\gamma,{\rm xn})$  experiment \cite{car71}.
The good agreement with real-photon induced experiments  speaks against specifics of the model-dependent conversion of Coulomb excitation to photoabsorption cross section or other experimental problems as a source of the discrepancy.   

The high energy-resolution achieved by dispersion matching enables investigations of the fine structure of the IVGDR with a wavelet analysis.
The comparison with RPA-based models with different degrees of sophistication provides an insight into the nature of characteristic wavelet scales related to the fine structure and in turn to competing mechanisms of giant resonance decay.
An in-depth study of $fp$-shell nuclei shows the unique presence of scales due to the coupling to more complex configurations (spreading width), but also an impact on scales due to Landau fragmentation.
Studies of the IVGDR in heavier-mass nuclei with the  (p,p$'$) reaction at RCNP using similar techniques demonstrate considerable fragmentation of the 1p-1h strength and corresponding wavelet scales up to $^{208}$Pb \cite{pol14,mar13,kru14}.
In deformed nuclei, a dominant source of fine structure is increased fragmentation of 1p-1h strength driven by $\alpha$ clustering in light-mass nuclei \cite{fea18} and $K$ splitting in heavy-mass nuclei \cite{don19}, as already observed for the ISGQR \cite{kur18}.
It should be noted that the IVGDR fine structure also allows the extraction of spin- and parity-separated level densities of $J^\pi = 1^-$ states which combined with corresponding high energy-resolution studies of other types of giant resonances \cite{kal06,vnc99,she04,she09} can provide important information on the spin \cite{egi09} and parity \cite{kal07} dependence of level densities.
The corresponding results from the present work will be discussed elsewhere.

To summarise, the present results demonstrate that inelastic proton scattering at 200 MeV and $0^\circ$ is a versatile tool for the study of the electric dipole response with relevance to nuclear structure, astrophysics and applications.
Further systematic studies including the even-mass Ca and Sm isotope chains are underway.

We are indebted to J.L.~Conradie, D.T.~Fourie and the accelerator staff at iThemba LABS for providing excellent proton beams. 
This work was supported by the National Research Foundation (South Africa) and by the Deutsche Forschungsgemeinschaft under contract SFB 1245. 

%

\end{document}